\RequirePackage{fix-cm}
\documentclass{svjour3}                     
\smartqed  
\usepackage{graphicx}
\usepackage{natbib,amssymb}
\usepackage{pdfpages} 
 \journalname{Experiments in Fluids}
\begin{document}

	\title{Analysis of soliton gas with large-scale video-based wave measurements}



	\author{I. Redor \and E. Barth\'elemy \and N. Mordant \and H. Michallet}


\institute{Laboratoire des Ecoulements G\'eophysiques et Industriels, Universit\'e Grenoble Alpes, CNRS,
Grenoble-INP, F-38000 Grenoble, France
}


\date{Received: date / Accepted: date}

\maketitle

\begin{abstract}
An experimental procedure for studying soliton gases in shallow water is devised. 
Nonlinear waves propagate at constant depth in a 34\,m-long wave flume. At one end of the flume, the waves are generated by a piston-type wave-maker. The opposite end is a vertical wall. 
Wave interactions are recorded with a video system using seven side-looking cameras with a pixel resolution of 1\,mm, covering 14\,m of the flume. 
The accuracy in the detection of the water surface elevation is shown to be better than 0.1 mm. 
A continuous monochromatic forcing can lead to a random state such as a soliton gas. 
The measured wave field is separated into right- and left-propagating waves in the Radon space and solitary pulses are identified as solitons of KdV or Rayleigh types.   
Both weak and strong interactions of solitons are detected.
These interactions induce phase shifts that constitute the seminal mechanism for disorganization and soliton gas formation.
\end{abstract}

\section{Introduction} 

A soliton gas is one of the possible random statistically stationary state of water wave motions. It is described by the theory of integrable turbulence \citep{zakharov1971kinetic,zakharov2009},
one of the theories along with the weak wave turbulence \citep{hasselmann1962,nazarenko2011} developed to describe the interaction of weakly nonlinear dispersive and random waves. 
Integrability of the equations exhibiting soliton solutions means that an infinite number of quantities amongst which mass, momentum, energy are conserved in time.
{It also implies that nonlinear coherent structures (such as solitons) emerge from any initial condition.}
A soliton gas is the statistical state in which solitons have random distributed amplitudes and phases.
Not only integrable turbulence poses theoretical
and experimental challenges in itself but is also one of the keys to understanding sea states generation and characteristics in shallow ocean basins \citep{costa}.
\\

Solitons are fascinating objects that propagate in weakly dispersive media. Since the first experimental observations of \cite{scott1844PRSE}
many other experiments have been reported in the literature \citep[e.g.][]{hammack1974,Seabra-Santos1987}. 
In the framework of water waves in shallow water, solitons are solutions of nonlinear equations such as 
the Korteweg-de Vries (KdV) equation \citep{Korteweg1895}. 
The KdV equation models a balance between non-linear and dispersive effects that shapes a solitary pulse of permanent form.
Nonlinear effects {steepen wave fronts while frequency dispersion stretches wave crests with components at different frequencies traveling with} different speeds.  
The free surface displacement due to a propagating soliton, also referred to as a solitary wave, is
\begin{equation}
\eta = a_s\ {\rm sech}^2\left[\frac{\beta}{2}\left(x-c \, t\right)\right]\,,
\label{solsech}
\end{equation}
where $\eta$ is the water surface elevation, $a_s$ the amplitude of the soliton, $c$ the phase speed, $\beta$ the shape factor, $x$ the space coordinate and $t$ time. 
The soliton is a localized symmetric hump traveling at constant speed. The parameters of this KdV solution fulfill
\begin{equation}
c = c_{\rm K}=\sqrt{g h}\left( 1+\frac{a_s}{2h} \right)
\label{ckdv}
\end{equation}
and
\begin{equation}
\beta=\beta_{\rm K}=\sqrt{\frac{3a_s}{h^3}}\,,
\label{betakdv}
\end{equation}
where $h$ is the quiescent water depth and $g$ the gravity acceleration. The $\beta$ parameter is a measure of the exponential decay of the trailing edges of the soliton.
The non-dimensional expression is
\begin{equation}
\beta^*_{\rm K}= \beta_{\rm K} \, h =\sqrt{\frac{3\,a_s}{h}}\,.
\label{betakdv_star}
\end{equation}
The shallow water KdV approximation assumes both {long waves and small amplitude}. 
Relaxing the latter restriction of small amplitude, an alternate solution known as the Rayleigh solitary wave solution is
\begin{equation}
c=c_{\rm R} =\sqrt{g \left( a_s + h \right)}
\label{crayleigh}
\end{equation}
and
\begin{equation}
\beta=\beta_{\rm R}=\sqrt{\frac{3 \, a_s}{h^2 \left( a_s + h \right)}} =  \frac{\beta_K}{\sqrt{1+ {a_s}/{h}}}\,.
\label{betarayleigh}
\end{equation}
The Rayleigh solitary waves decay more gradually in the trailing edges than those of KdV: they are wider.
\\

The experimental generation of soliton gases is challenging. These random states emerge on very long time scales and thus require long-term experiments.
Very recently \citet{redorPRL}, drawing on the fission of sine-type long waves into multiple solitons \citep{zabusky1971,Trillo}, 
used a continuous wave-maker forcing in order to reach a soliton gas state, such a state only obtained numerically so far \citep[e.g.][]{Pelinovsky,carbone}.
The wave generation set-up is similar to that used by \cite{Ezersky2} but in a longer flume in order to favor sine-wave complete fission into solitons. 
A piston-type wave maker is driven with a small regular motion at one end and the waves are free to reflect against a vertical wall at the other end. 
Reflection of solitons, dynamically equivalent to head-on collisions \citep{Renouard1985,Chen2014}, 
that also occur in the experiments produce soliton amplitude reduction, dispersive wave generation and phase shifting {\citep[see also][for a thorough analysis of the flow field]{chen2014laboratory2}}. Strong interaction (overtaking collision) also induces a phase shift to the solitons {\citep[e.g.][]{umeyama2017}}.
These many collisions of the two types concur to the emergence of a bi-directional soliton gas. Soliton amplitude random distribution is also a result of the dissipative decay due to friction
as modeled by \citet{keulegan1948} with a boundary layer model.
\\

The randomness and bi-directionality of the wave field makes it difficult to track the solitons traveling
back and forth. Even though arrays of closely arranged wave probes and repeated experiments to cover large regions of interest \citep[e.g.][]{Taklo2017} are possible,
actual improvements in video resolution provide image based measurements with sufficient spatial resolution
to separate waves traveling in both directions. Video techniques such as slope detection and relative brightness for statistical properties 
of wave fields \citep{zhang1994,chou2004}, bichromatic synthetic Schlieren \citep{kolaas2018}, free surface synthetic Schlieren \citep{moisy2009} and
Fourier transform profilometry \citep{cobelli2009,aubourg2016}, are widely used to measure 2D water wave motions.
\cite{bonmarin1989} used a side-looking CCD camera to image the interface in breaking waves in a 1D wave flume. 
The technique relies on tinted water and sharp lighting to create strong contrasts at the interface amenable to edge detection processing.
Although the foundations of this technique with coarse resolution are close to those described in this paper, it was only used to measure one wave length. Recent technologies allow for multiple synchronized camera settings
and large redundant storage capacities as presented hereafter.
The innovative system used for this study images a $14$-m-long section of a $34$-m-long flume with sub-millimetric resolution, enabling to capture highly nonlinear wave interactions in random states.
In addition, we use the Radon transform \citep{deans2007} to separate, in the space-time field, the direction of propagation of the solitons and to determine their phase speed.
As shown by \cite{Sanchis2011} for instance, the Radon transform is well suited to the detection of lines in noisy images. 
\\

The experimental set-up is described in detail in \citet{redor2019}, we recall the main characteristics in section~\ref{setup}.
Measurement accuracy, wave separation methods, analysis of soliton head-on collision and soliton dissipation, are presented in section 3.
The soliton gas experiments are described and analyzed in section~\ref{gaz}.
Section 5 is devoted to the conclusions.

\section{Experimental set-up} 
\label{setup}

\begin{figure}
\centerline{
\includegraphics[width=75mm]{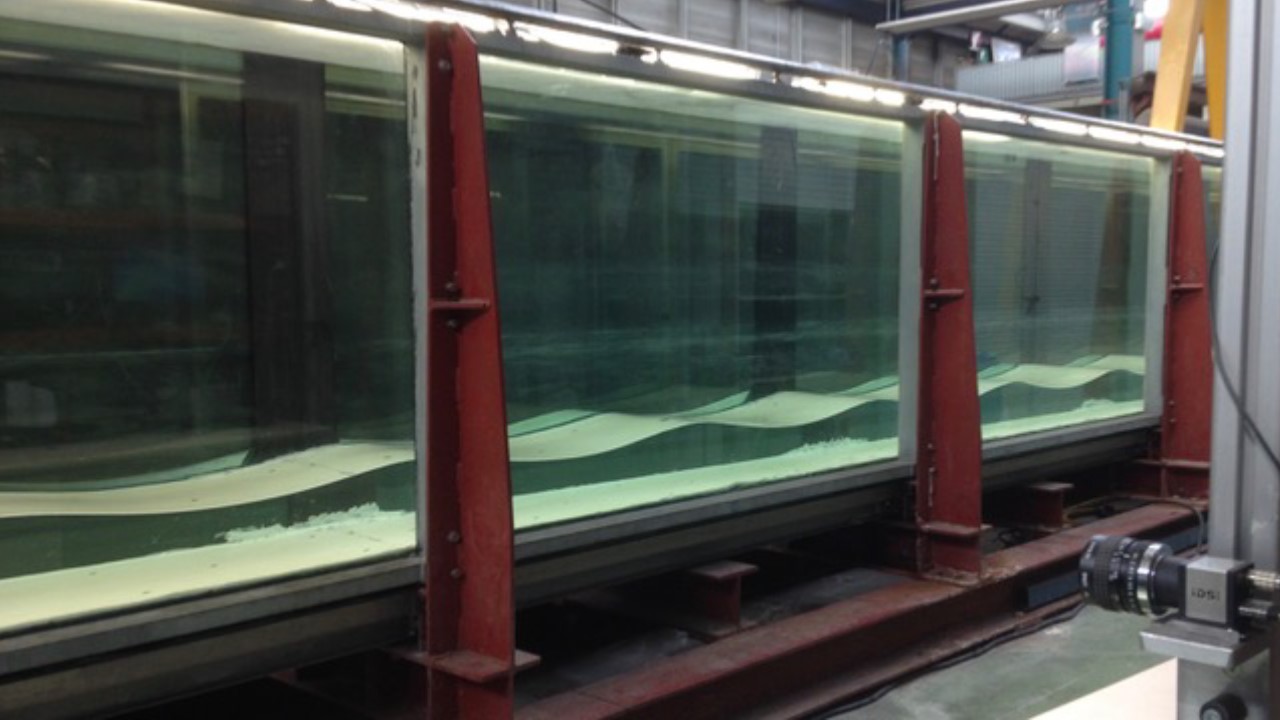}}
\caption{View of a section of the flume, with one of the cameras (bottom right corner). 
Each camera records the water surface elevation along the 1.92\,m separating two vertical beams (red-brownish color).
\label{canal}}
\end{figure}

\subsection{Wave flume} 

The flume is horizontal within a few millimeters over its entire length $L=33.73$ m. Its width is $l=0.55$ m.
The flume is made of lateral glass windows of 1.92-m-long maintained together by steel cradles with vertical beams 8-cm-wide (see Figure~\ref{canal}).

The left-hand boundary of the flume is a movable vertical piston-type wave-maker of 0.6\,m maximum stroke.
It is controlled in position, with an accuracy of approximately 1 mm, to produce any prescribed motion of limited acceleration (roughly frequencies $f\leq 2$ Hz).
Following \citet{Guizien2}, solitons with minimum trailing dispersive tail can be generated according to Rayleigh's law of motion, i.e. a single forward push of tanh shape. 
In the present paper, we consider interactions between solitons. Producing several solitons with Rayleigh's law implies to move back the piston before each new forward push. 
To overcome this limitation, solitons can be produced through nonlinear steepening of a regular monochromatic forcing of large enough amplitude, as shown in section~\ref{gaz}.

\subsection{Video } 

\begin{figure}
\centerline{
\includegraphics[width=125mm]{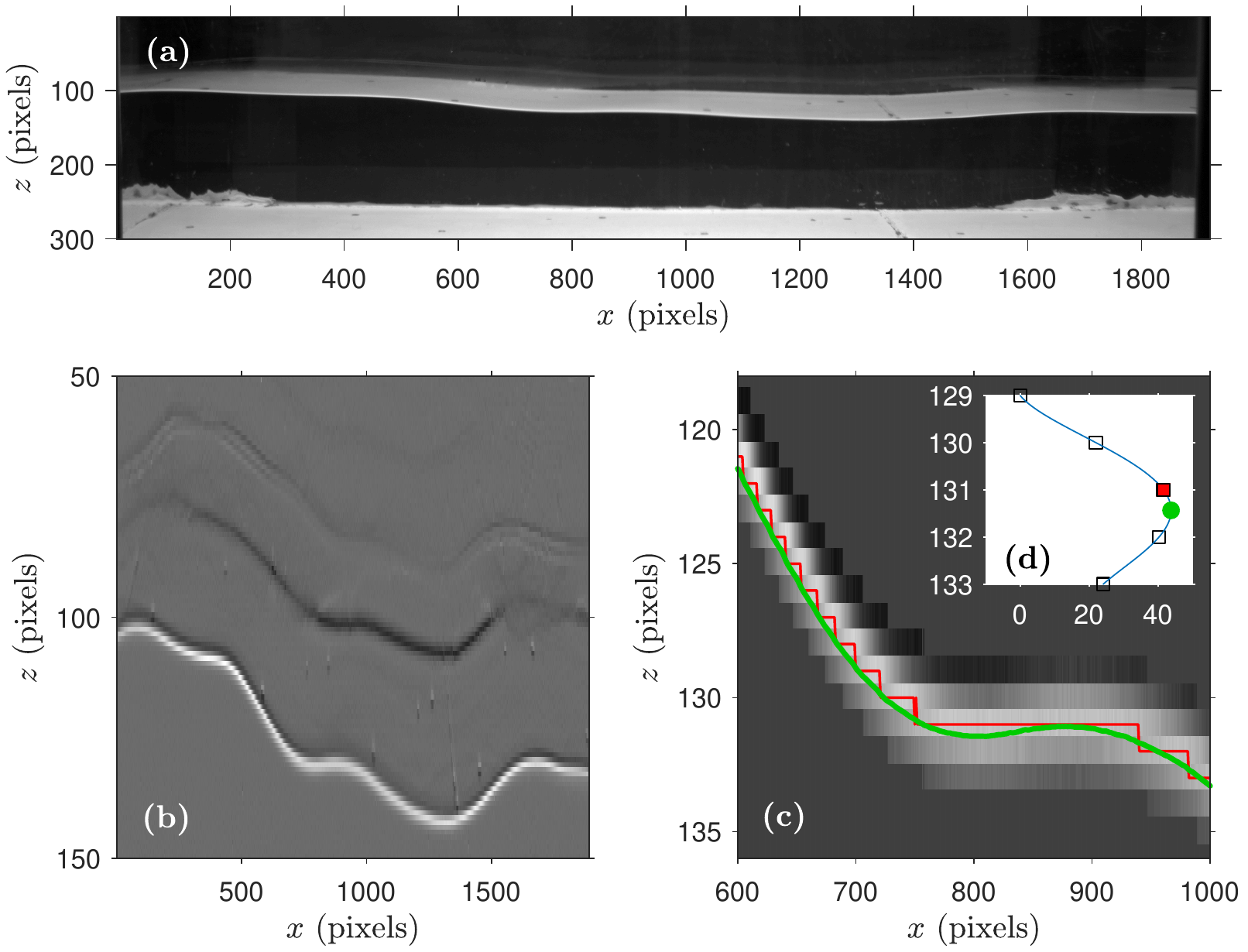}}
\caption{
(a) Snapshot of a camera field of view in $2^8$ gray levels. The bright regions are the bottom of the flume and the water surface, seen with a slight down-looking angle.
(b) Region of interest in gray level gradient.
(c) Close-up with detected maximum gradient (red line) and interpolated interface (green line).
(d) Vertical profile, at $x=800$ pixels, of the gray level gradient (squares), with  maximum (filled square), $4^{\rm th}$ order polynomial best fit (solid line) and interpolated interface position (green dot).
\label{interf}}
\end{figure}

In most experiments, the water elevation is recorded at 20~frames/s (maximum 100 fps) with seven cameras located 2 m aside of the flume (see Figure~\ref{canal})  covering seven glass windows 1.92 m long.
The flume bottom is painted white and the background wall is black. The flume is illuminated from above the free surface. 
Each camera is slightly downward-looking the water surface. This provides the best contrast for following the water surface along the front glass wall.
An example of a raw image is shown in Figure~\ref{interf}(a). 
The image size is $1936\times 300$ pixels, so that the pixel resolution is about one square millimeter. 
Pictures of a regular grid are used for calibration \citep[see e.g.][]{tsai1987versatile}.

The region of interest is automatically detected {to remove} the flume beams and flume bottom. 
A first estimation of the interface position is the line of pixels of maximum gray level gradient, see Figure~\ref{interf}(b). 
Sub-pixel resolution is achieved by fitting a
$4^{\rm th}$ order polynomial curve in the gray level gradient over a vertical column of five pixels, see Figure~\ref{interf}(d).  
The maximum of that curve provides the interpolated interface position, see Figure~\ref{interf}(c). 

Many other proxies for interface position have been tested, such as darkest or brightest pixels. 
The method described  above  appears to be the best for minimizing bias or hysteresis related to glass wall capillary meniscus, as will be shown in the following.

\section{Wave analysis} 

\subsection{Water drop experiment} 
\label{drop}

\begin{figure}
\centerline{
\includegraphics[width=150mm]{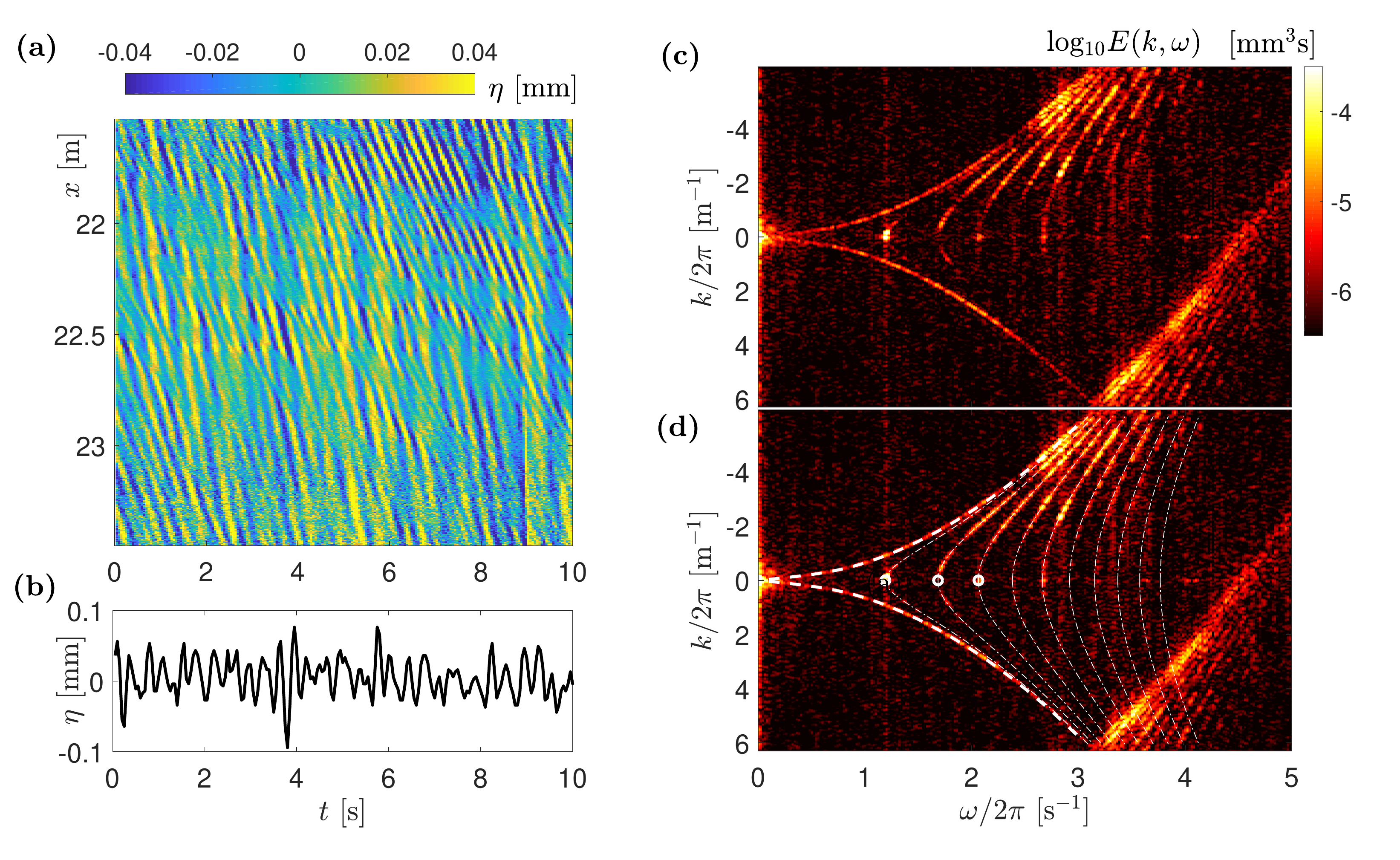}}
\caption{The water drop experiment.
(a) Time-space representation of the free surface elevation in the field of view of the last camera (drop injection is at $x=0$).
(b) Time series extracted from (a) at $x=22.5$ m.
(c) $E(k,\omega)$ spectrum computed from the full field of view ($9.53 \leq x \leq 23.45$ m) with a 0.08 m resolution.
(d) Same as (c) with superimposed dispersion relations : 
linear waves propagating along $x$ (dashed) and combined waves (dotted) related to resonant transverse modes (the three first ones are marked with circles).
\label{fdrop}}
\end{figure}

In order to assess the accuracy of the free surface detection procedure and measurement, the following experiment is conducted. 
The flume is filled at a water depth of $h=0.7$ m. 
At one end of the flume a small vertical tube drips, releasing water drops {that impact the water surface every 5 to 10\,s}.
The waves generated by the successive impacts are recorded with the cameras. An example of the free surface elevation record is shown in Figure~\ref{fdrop}(b), 
{for which the standard deviation of the free surface displacement is a few hundredths of millimeter}. 
In the space-time field in Figure~\ref{fdrop}({a}) the color ridges are the signature of waves  crests and troughs that essentially propagate from {left to right} in the flume. 

The 2D Fourier spectrum shown in Figure \ref{fdrop}(c) for this experiment is computed over the full 14-m-long field of view as the square modulus of the time-space Fourier transform $\tilde\eta(k,\omega)$:
\begin{equation}
E(k,\omega)={\left| \int\!\!\!\!\int \eta(x,t) e^{i (k x+\omega t )} \mathrm{d} x \mathrm{d} t \right|}^2\,. \label{rel:FFT}
\end{equation}
For $\omega>0$, negative wave numbers $k<0$ correspond to waves propagating 
with increasing $x$ (similarly for $\omega < 0$ and $k > 0$ as the field is real), while $k>0$ represent waves reflected by the flume end-wall for $0<\omega/2\pi<3$ s$^{-1}$.  
Due to spectral aliasing (in wave number) for this spatial resolution of 8 cm, the energy for $k>0$ and $\omega/2\pi>3$ s$^{-1}$ {corresponds} to waves propagating towards the end-wall.  

The energy in the ($k,\omega$) plane is mostly located along curves, see Figure \ref{fdrop}(d). 
On the one hand waves propagating along the $x$ axis in both directions match the dispersive relation of linear waves:
\begin{equation}
\omega=\pm (gk\tanh(kh))^{1/2}  \,.
\label{eqdisp}
\end{equation}
On the other hand the wave numbers of resonant transverse modes with anti-nodes at the lateral walls match
\begin{equation}
 k_y=\frac{\pi n}{l}\quad{\rm with}\quad n\in\mathbb{N} \,.
\end{equation}
These transverse waves correspond to dots along the $k=0$ axis with $\omega=(gk_y\tanh(k_yh))^{1/2}$.
Their interactions with waves propagating along $x$ produce transverse waves that comply the following dispersive relation:
\begin{equation}
\omega=\left(         g\left(k^2+  {k_y}^2\right)^{1/2} \tanh\left( \left(k^2+  {k_y}^2\right)^{1/2}h\right)       \right)^{1/2} \,.
\end{equation}

The good agreement of the measurements with these theoretical estimators indicates that the accuracy of the video sub-pixel detection can be of the order of a few hundredths of a millimeter.
\\

\subsection{Wave separation} 
\label{wavesepa}

\begin{figure}
\centerline
{{\includegraphics[height=50mm,viewport=210 180 540 360,clip]{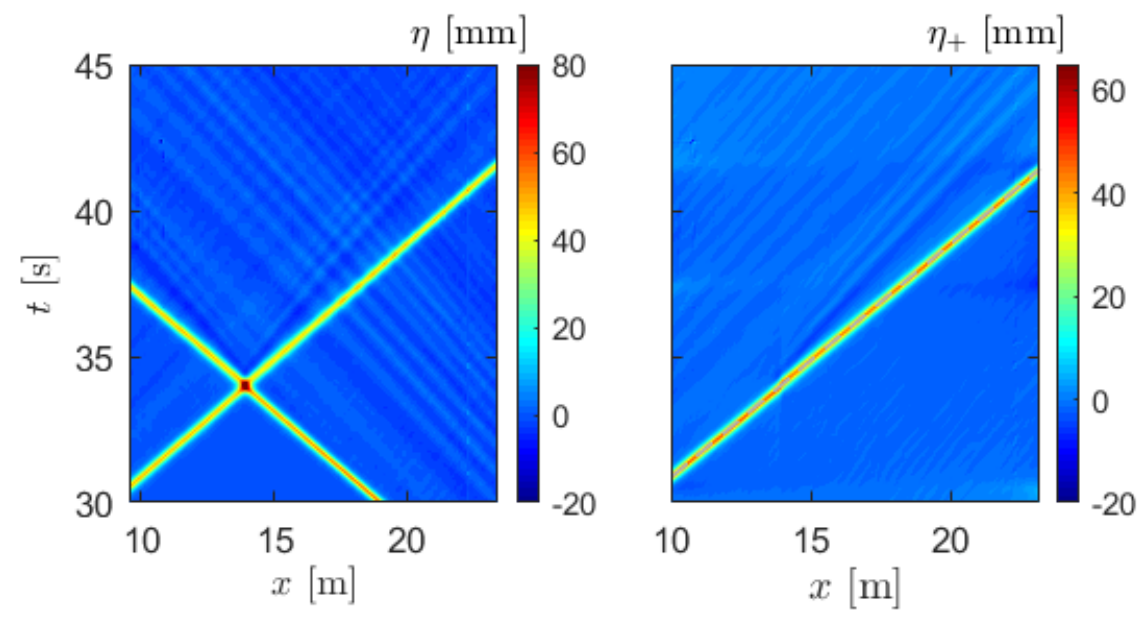}}
{\includegraphics[height=50mm,viewport=20 180 200 360,clip]{fft2D_13_34.pdf}}}
\vspace*{-5.2cm}
\hspace{0.8cm} {\bf (a)} \hspace{4.7cm} {\bf (b)} \hspace{3.8cm} {\bf (c)}\\[4.5cm]
\caption{Head-on collision of two solitons: water surface elevation (a) and separated right-running waves (b), 
time-space Fourier spectrum (c): in dashed is the linear dispersion relation (\ref{eqdisp}), in solid is the relation $\omega=\pm c_0 k$. 
Water depth: $h=12$ cm, right-running soliton amplitude: $a_+=4.7$ cm, left-running soliton amplitude: $a_-=5.0$ cm. 
\label{fftchpintfaible}}
\end{figure}

\subsubsection{Fourier transform}

In order to get a better understanding of wave interactions in the bidirectional flow,  waves propagating towards positive $x$ ('right-running') 
can be separated from those propagating towards negative $x$ ('left-running') by computing the inverse Fourier transform of selected {quadrants} of the time-space spectrum computed with (\ref{rel:FFT}).
An example is shown in Figure \ref{fftchpintfaible}.
Figure \ref{fftchpintfaible}(c) is the time-space spectrum of the interaction of two solitons propagating in opposite directions. 
The time-space field of surface elevation is plotted in Figure \ref{fftchpintfaible}(a). The signatures of the two solitons are identified as the bright straight lines. 
Waves with $k < 0$ and $\omega > 0$ propagate towards positive $x$ and waves with $k > 0$ and $\omega > 0$ to negative $x$. 
The energy in Figure \ref{fftchpintfaible}(c) is mostly located along straight lines, which confirms that all the frequency components of each soliton propagate at the same speed
$c=\left|\omega/k\right|>c_0$, where $c_0=\sqrt{g h}$ 
is the linear long wave phase speed. 
Weaker energy levels are located along the dispersion relation of linear waves (\ref{eqdisp}) {which is the} signature of weak dispersive waves. 
Applying the inverse Fourier transform on $\tilde\eta(k<0, \omega>0)$ produces the field shown in Figure \ref{fftchpintfaible}(b). 
Small amplitude dispersive waves are visible in the top-left corner of the plot, corresponding to waves generated in the lee of the soliton by the wave-maker motion. 
Additional small waves are generated at the collision.

\subsubsection{Radon transform}
\label{radon}

An alternative to the Fourier transform for separating right- and left-running waves is the Radon transform \citep{deans2007,Almar2014}. 
The Radon transform is the projection of a field intensity along a radial line oriented at a specific angle $\theta$:
\begin{equation}
R(\rho,\theta) = \int\!\!\!\!\int \eta(x,t) \, \delta\!\left({\frac{x}{\Delta x}}\cos\theta+{\frac{t}{\Delta t}}\sin\theta-\rho\right) \, \mathrm{d} x \, \mathrm{d} t
\end{equation}
where $\delta$ is the Dirac function and  
{$\rho=((x/\Delta x)^2+(t/\Delta t)^2)^{1/2}$
is the distance (in pixels, with $\Delta x$ and $\Delta t$ the spatial and temporal resolutions, respectively)}
from origin (center of the two-dimensional field).

\begin{figure}
\centerline
{{\includegraphics[height=50mm,viewport=50 180 240 360,clip]{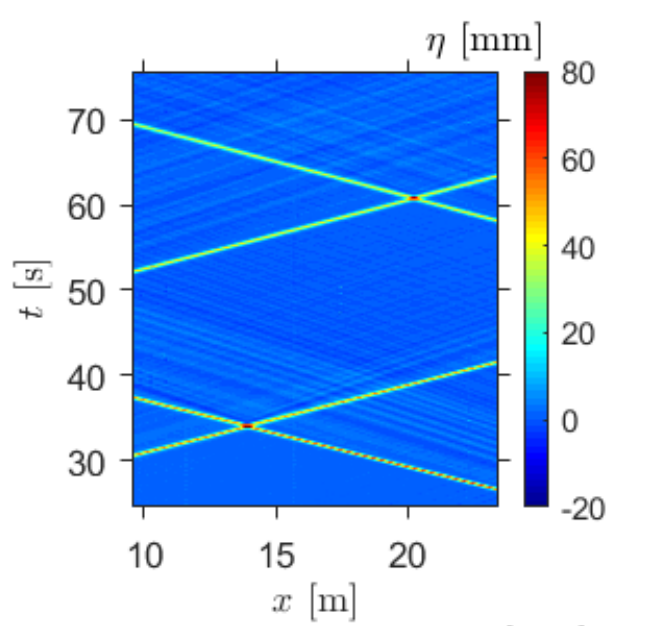}}
{\includegraphics[height=50mm,viewport=280 180 430 360,clip]{cradon_13_34s_14m_h.pdf}}
{\includegraphics[height=50mm,viewport=430 180 610 360,clip]{cradon_13_34s_14m_h.pdf}}}
\vspace*{-5.1cm}
\hspace{0.8cm} {\bf (a)} \hspace{4.7cm} {\bf (b)} \hspace{3.8cm} {\bf (c)}\\[4.5cm]
\caption{Two head-on collisions of two solitons (same experiment as in Figure~\ref{fftchpintfaible}): (a) water surface elevation and (b) separated left-running waves, (c) projection of $\eta(x,t)$ in the Radon space $R(\theta,\rho)$. 
The angle $\theta$ is the inclination relatively to the vertical axis in the $(x,t)$ plane, $\rho$ is the distance to the center of the field $\eta(x,t)$, 
as drawn in (b) for the local maximum marked with a circle in (c). The vertical lines are angle values $\theta_0=69.8^\circ$ and $110.2^\circ$ that both correspond to $c_0 =1.07$ m s$^{-1}$ in the physical space $(x,t)$.
\label{chpintfaible}}
\end{figure}

For example, we consider the $\eta(x,t)$ field shown in Figure \ref{chpintfaible}(a) of two solitons passing twice in front of the cameras. 
Four local maxima are detected in the Radon space shown in Figure \ref{chpintfaible}(c). 
These four maxima correspond to the four crest trajectories of the $\eta(x,t)$ field in Figure \ref{chpintfaible}(a). The local maxima $\theta$ values correspond to the propagation speed of the four solitons. 
Applying the inverse Radon transform on $R_- = R(\rho,\theta>90^\circ)$ produces the plot in Figure \ref{chpintfaible}(b) where only left-running waves are reconstructed. 
Note that the local maxima $\theta$ values for the left-running waves are below the $\theta_0$ associated to $c_0$, confirming that the solitons travel faster than $c_0$.
\\

\begin{figure}
\centerline
{{\includegraphics[height=50mm,viewport=430 180 617 360,clip]{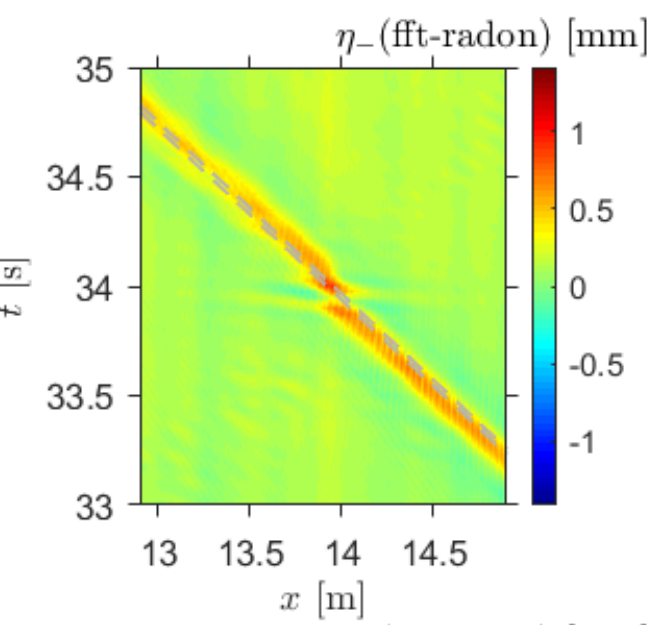}
{\includegraphics[height=50mm,viewport=275 180 610 360,clip]{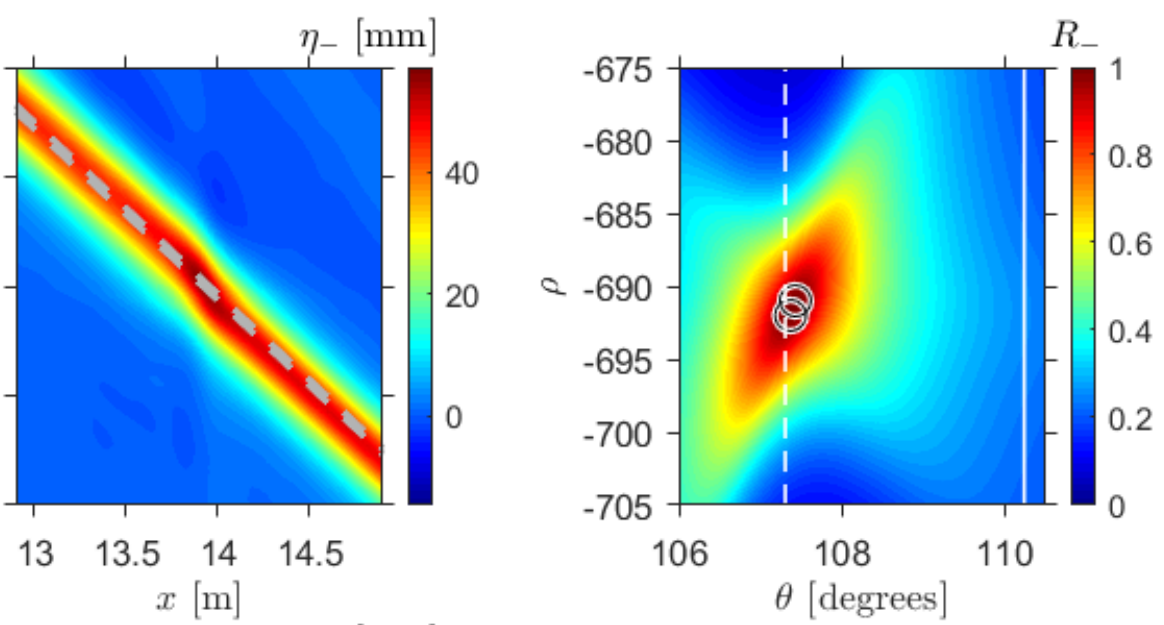}}
}}
\vspace*{-5.2cm}
\hspace{0.8cm} {\bf (a)} \hspace{4.7cm} {\bf (b)} \hspace{3.8cm} {\bf (c)}\\[4.6cm]
\caption{Close-up of the left-running soliton in the head-on collision shown in Figure \ref{chpintfaible}(a) at ($x=14$ m, $t=34$~s): (a) difference between separated  left-running wave fields from Fourier and Radon, 
(b)  left-running wave, (c) corresponding signature in the Radon space $R_-(\theta,\rho)$. 
The vertical solid line corresponds to $c_0$, the vertical dashed line corresponds to the speed (\ref{crayleigh}) of a Rayleigh soliton of amplitude $a=0.4 h$. 
Two detected local maxima are  marked with circles, their transposition into the physical space $(x,t)$ are marked as dashed lines in (a,b) and correspond to the soliton before and after interaction. 
\label{chpintfaible2}}
\end{figure}

Fourier and Radon separation methods give similar results. 
For instance, we compare the reconstructed fields of the soliton head-on collision. 
The difference between left-running reconstructed fields is shown in Figure \ref{chpintfaible2}(a) to be less than 1\,mm.
The focus on this left-running separated wave shown in Figure \ref{chpintfaible2}(b) highlights the phase lag that a soliton undergoes during a head-on collision. 
Indeed the color ridges before and after the collision are not aligned.
Moreover both separation methods assign a part of the overall amplitude amplification due to non-linearity ($\eta_{\max}>a_++a_-$, with $a_+$ and $a_-$ the amplitudes of the right- and left-running solitons, respectively) 
to each of the solitons, as will be discussed in more detail below in section~\ref{sweakint}.
\\

The corresponding signature in the Radon space $R_-(\rho,\theta>90^\circ)$, in Figure \ref{chpintfaible2}(c), exhibits two local maxima that correspond to waves propagating 
at a roughly same constant speed 
in the physical plane, {emphasized} by dashed lines in Figure \ref{chpintfaible2}(b). 
They are associated to the soliton trajectory before and after the head-on collision.
Their speeds are very close to that of a KdV or Rayleigh soliton of amplitude $a=0.4 \, h$.
This example emphasizes that the Radon transform is also a tool for identifying soliton trajectories, their speeds and eventually phase shifts.
\\

\begin{figure}
\centerline
{{\includegraphics[width=125mm,viewport=250 100 650 480,clip]{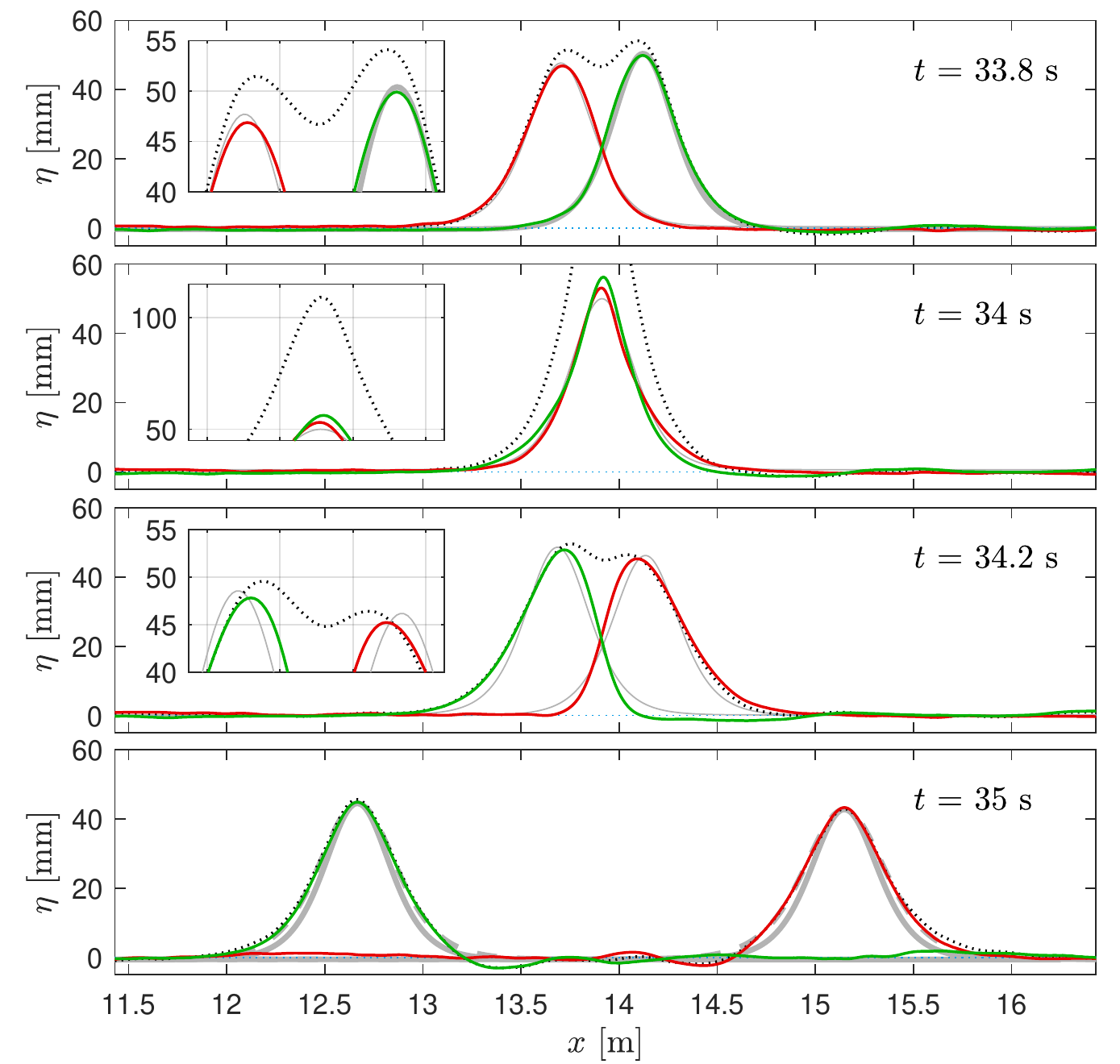}}}
\caption{Head-on collision of two solitons: water surface elevation (dotted) separated into right-running soliton (red) and left-running soliton (green). 
Gray solid lines are KdV profiles matching the measures at the beginning of the collision (top panel), gray dashed lines are Rayleigh profiles matching the {measurements} 1~s after collision (bottom panel). 
During the collision (second panel) note a nonlinear amplification resulting in a maximum free surface elevation exceeding the linear superposition of the two incident waves, as shown by the inserts. 
The interaction is inelastic, characterized by asymmetric wave profiles at the end of the collision (third panel) and the generation of dispersive waves in the lee of each soliton. 
\label{solintfaible}}
\end{figure}

\subsection{Weak interaction of two solitons} 
\label{sweakint}

Free surface elevation time series of the head-on collision of the experiment of Figure \ref{chpintfaible} are shown in Figure \ref{solintfaible}.
The measured profiles are in fair agreement with theoretical KdV profiles (\ref{betakdv}) before and long after the collision. 
During the collision the separated solitons are stretched, so that the measured maximum surface elevation exceeds that of the sum of the incoming soliton amplitudes. 
In this example, the conditions ($a_+/h= a_-/h= 0.4$) are similar to that of \cite{Chen2014} and the maximum amplification is $\eta_{\max}/(a_++a_-)=1.13$ too. 
Just after the collision, before the dispersive tails {detach} from the solitons, the measured wave profiles tally more closely to the Rayleigh soliton profiles (\ref{betarayleigh}).
\\

\begin{figure}
\centerline
{{\includegraphics[width=125mm,viewport=250 55 650 320,clip]{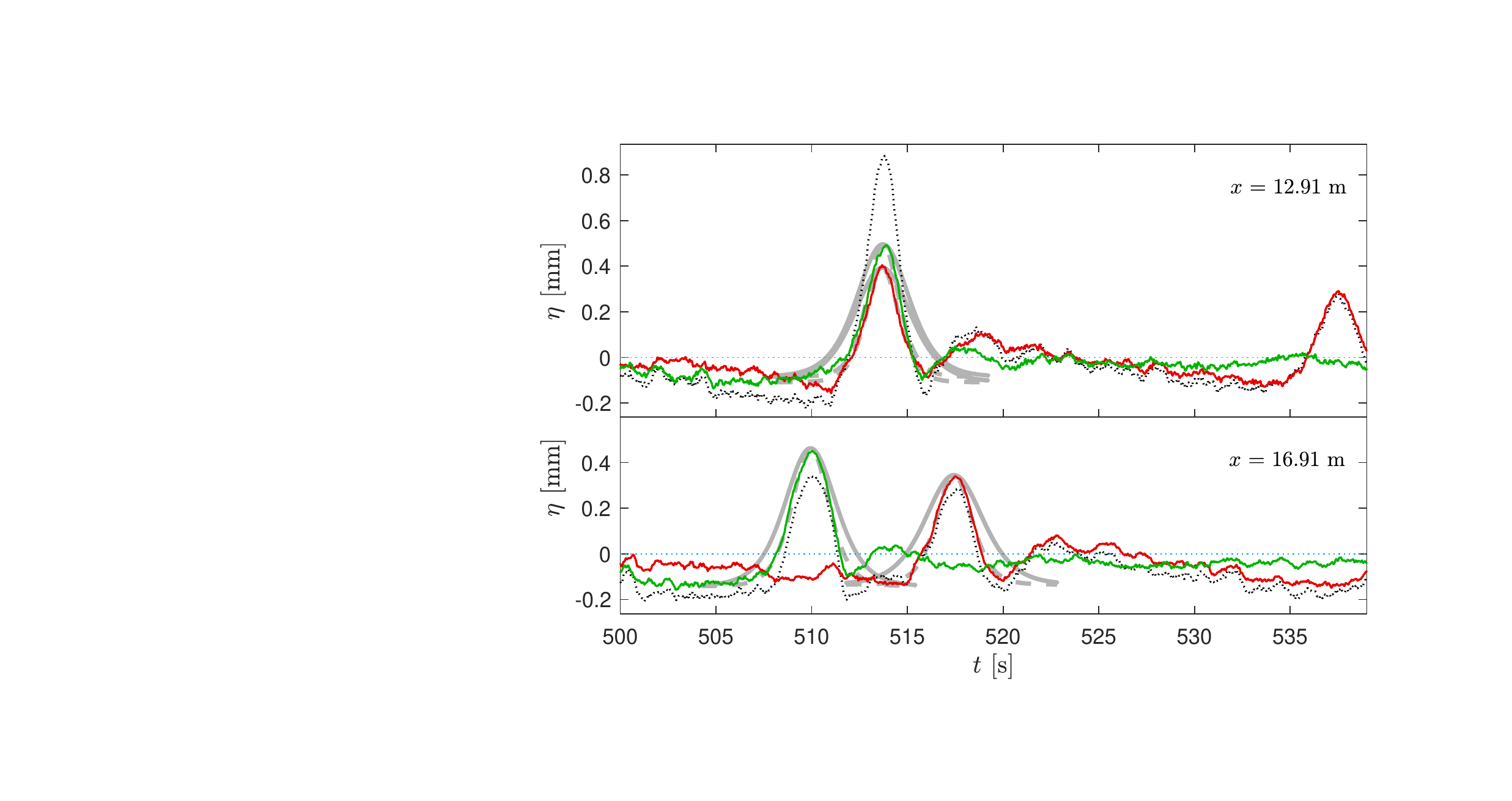}}}
\caption{Time series of water surface elevation (dotted) separated into right-running wave (red) and left-running wave (green). Gray solid lines are KdV profiles, gray dashed are sech$^2$ fitted profiles. 
Same experiment as in Figure~\ref{solintfaible}.}
\label{solintfaible2}
\end{figure}

In the same experiment, the two solitons continue to propagate back and forth in the flume, reflecting at both ends and interacting. 
Even for large propagation durations solitary waves pulses matching a sech$^2$ profile (\ref{solsech}) are still detected (see Figure~\ref{solintfaible2}).
The solitary waves in Figure~\ref{solintfaible2} have traveled 7 times back and forth and the waves are a few tenths of a mm high. 
The dashed gray lines represent solitons (\ref{solsech}) with 3 fitted parameters $a$, $\beta$ and $h_r$, the latter parameter $h_r$ accounting for a soliton propagation at a reference level 
different than $h$.
The fitted shape factor $\beta$ is here larger than that of the KdV solution (\ref{betakdv}), meaning that the measured solitons are narrower. 
This is probably due to the weak background flow induced by the wave-maker backward slow motion (after each forward push that produces a soliton) that generate a standing seiching wave. 
At long times, each soliton is leading a remnant long wave crest propagating at $c_0$ too. This also corresponds to $h_r<h$ of about one tenth of a mm just in front of each leading soliton in Figure~\ref{solintfaible2}.
\\

All the elements discussed in this section indicate that our video system can accurately capture nonlinear wave dynamics.
Automatic detection of wave crests and their propagation speed with the Radon transform is a tool to determine soliton characteristics in a wave field.

\subsection{Adiabatic soliton propagation} 
\label{adia}

\begin{figure}
{
\hspace*{1.3cm} {\bf (a)} \hspace{6.6cm} {\bf (c)} \\
\hspace{.7cm}
{\includegraphics[width=65mm,viewport=60 30 275 330,clip]{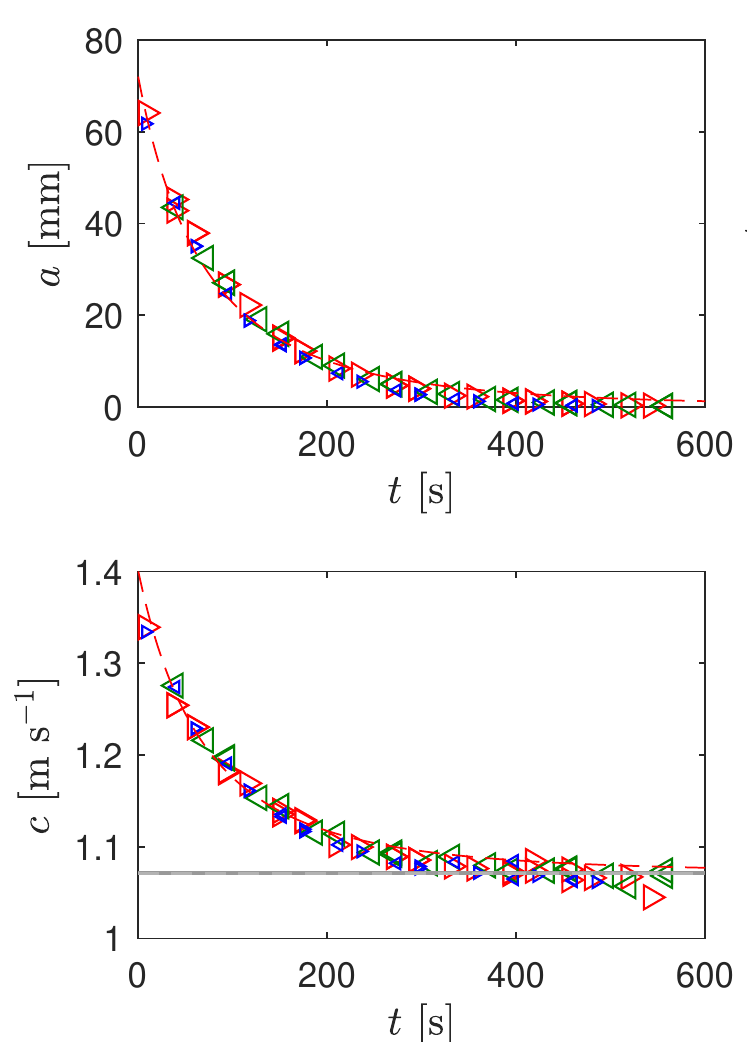}}
\hspace{1.5cm}
{{\includegraphics[width=65mm,viewport=490 30 705 330,clip]{Cbeta_21_17e.pdf}}}\\
\hspace*{1.3cm} {\bf (b)} \hspace{6.6cm} {\bf (d)} \\
}
\caption{ 
Time variations of crest (a) maximum and (b) speed detected at $x=17$~m.
The horizontal line is $c_0 = 1.07$ m s$^{-1}$, the red dashed lines represent Keulegan's dissipation decay laws. 
(c) Dimensionless speed against dimensionless amplitude and
(d) dimensionless shape factor $\beta_*= \beta h c/c_0$ against dimensionless soliton amplitude.
Thick solid and dashed lines are for KdV and Rayleigh solutions, respectively.
Right (left) pointing triangles are for right (left) propagating waves. 
Small triangles correspond to a single soliton traveling back and forth in the flume. Large triangles correspond to the head-on collision experiment.  
\label{soldissip}}
\end{figure}

The time evolution of soliton amplitude and speed is plotted in Figure~\ref{soldissip}(a, b), for the head-on collision experiment described above 
as well as for a single soliton propagating back and forth in the flume.
Both amplitude and speed decays show good agreement with the boundary layer dissipation decay laws proposed by \citet{keulegan1948}.
\\

The soliton profiles match well nonlinear theoretical solutions, with best fit parameters lying in between KdV and Rayleigh solutions, 
as measured at the middle of the flume and plotted in Figure \ref{soldissip}(c,d) for instance.
\\

Despite dissipation, the solitons propagating back and forth in the flume closely fulfill at any time a balance between dispersive and non-linear effects. 
In that sense the soliton propagation can be considered adiabatic, and thus integrable. 

\newpage
\section{Soliton gas} 
\label{gaz}

\begin{figure}
\hspace*{.3cm} {\bf (a)} \hspace{4.6cm} {\bf (b)} \hspace{4.6cm} {\bf (c)}\\
{{\includegraphics[height=43mm,viewport=40 183 247 358,clip]{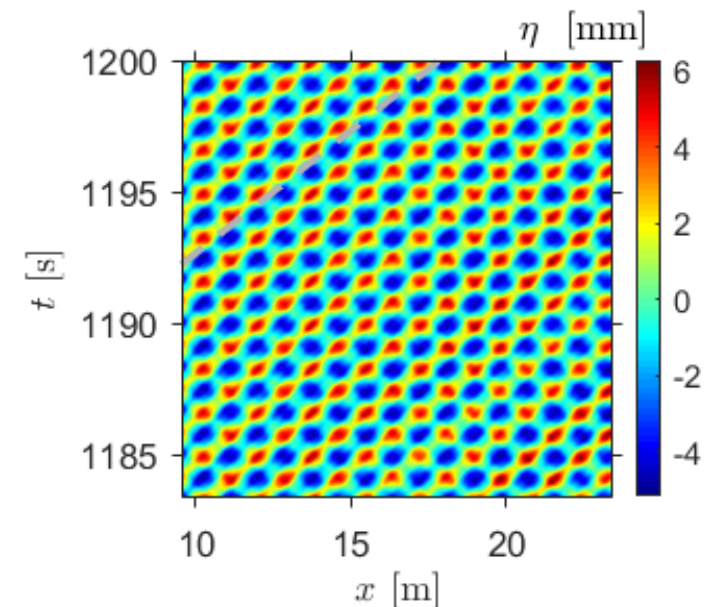}}}
{{\includegraphics[height=43mm,viewport=40 183 247 358,clip]{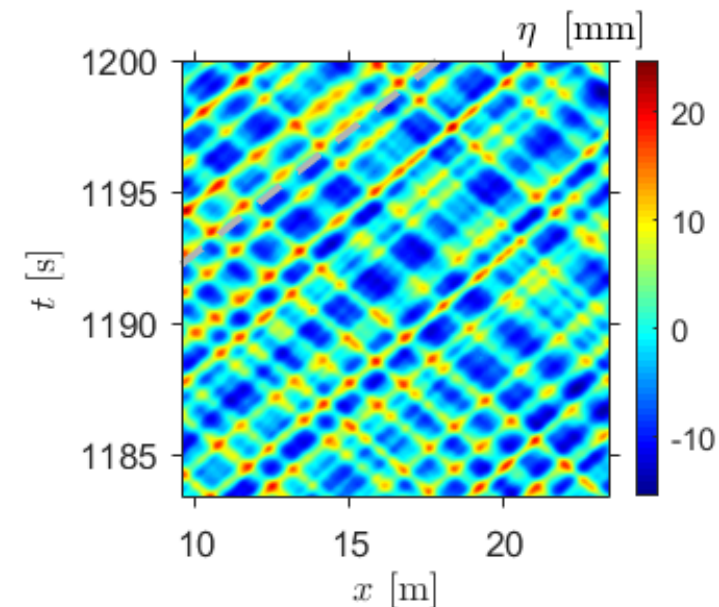}}}
{{\includegraphics[height=43mm,viewport=40 183 247 358,clip]{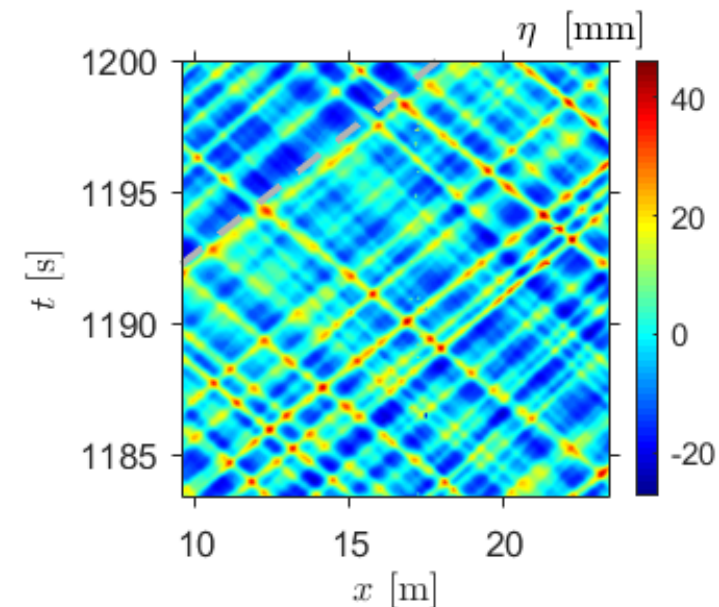}}}\\
\hspace*{.3cm} {\bf (d)} \hspace{4.6cm} {\bf (e)} \hspace{4.6cm} {\bf (f)}\\
{{\includegraphics[height=44mm,viewport=245 183 448 358,clip]{eta_x_t_16_1200n.pdf}}}
{{\includegraphics[height=44mm,viewport=245 183 448 358,clip]{eta_x_t_10_1200n.pdf}}}
{{\includegraphics[height=44mm,viewport=245 183 448 358,clip]{eta_x_t_11_1200n.pdf}}}\\
\hspace*{.3cm} {\bf (g)} \hspace{4.6cm} {\bf (h)} \hspace{4.6cm} {\bf (i)}\\
{{\includegraphics[height=44mm,viewport=245 5 448 180,clip]{eta_x_t_16_1200n.pdf}}}
{{\includegraphics[height=44mm,viewport=245 5 448 180,clip]{eta_x_t_10_1200n.pdf}}}
{{\includegraphics[height=44mm,viewport=245 5 448 180,clip]{eta_x_t_11_1200n.pdf}}}
\caption{ 
Space-time surface elevation (a-c), space-time Fourier spectrum after 20 min of forcing (d-f) with solid lines indicating $\omega=c_0 \left|k\right|$ and dashed lines for the dispersion relation (\ref{eqdisp}), 
and time variation of the spatial spectrum (g-i). 
Water depth $h=12\,$cm ($c_0 = 1.07 \,$m/s), sinusoidal forcing with frequency $\omega/2\pi=0.6$ s$^{-1}$ and amplitude $a_0=4$ mm (a,d,g), $a_0=6$ mm (b,e,h), $a_0=12$ mm (c,f,i).
\label{gas}}
\end{figure}

In the previous section the measuring system {was} validated along with the processing tools adapted to nonlinear wave propagation.
Hereafter we describe and analyze, in particular on large time scales, the wave field generated by a continuously right-running sinusoidal wave.
The flume water depth is $h=12$\,cm, the sinusoidal wave surface elevation is 
$\eta(x=0, t)=a_0\sin(\omega_0 t)$, with $\omega_0/2\pi=0.6$\,s$^{-1}$.
Three different forcing amplitudes are considered: 
$a_0 = 4$, 6 and 12\,mm. 
The waves are free to reflect at both ends of the flume.
\\

Figure~\ref{gas}(a-c) shows  space-time representations of the  wave field after 20 minutes for the three cases.
For the smallest amplitude plotted in Figure~\ref{gas}(a), a steady state is reached exhibiting a standing wave pattern. 
This is confirmed by the space-time spectrum in Figure~\ref{gas}(d) where the energy is mainly located at the forcing frequency and its harmonics $2\omega$ and $3\omega$. 
Figure~\ref{gas}(g) is the time variation of the spatial spectrum for the same experiment, indicating that the steady state is rapidly reached, say after about 200~s 
or a few flume-length travels of the first wave train. 
Of note, {the two red ridges} for $10<t<100$~s for $k/2\pi=1.15$~m$^{-1}$ and $1.26$~m$^{-1}$ corresponding to both bound and free second harmonic, respectively. 
\\

For a slightly larger forcing ($a=6$ mm), the evolution is very similar during the first 100\,s, see Figure~\ref{gas}(h). 
At later times, energy is distributed over a wider range of wave numbers. 
After 20 minutes of forcing, two red spots at the main frequency $\omega_0/2\pi=0.6$\,s$^{-1}$and $k_0/2\pi=\pm 0.58$~m$^{-1}$ are clearly visible in Figure~\ref{gas}(e), 
but  energy located along the straight lines $\omega/k=\pm c_0$ denotes the presence of solitons. 
The corresponding wave field (Figure~\ref{gas}b) is characterized by bright ridges that are the signature of solitons propagating at various speeds. 
\\

These features are even more obvious for the larger forcing amplitude $a=12$~mm in Figure~\ref{gas}(c, f, i). 
Energy is distributed over a large frequency band, see Figure~\ref{gas}(f), with no dominance of the forcing frequency after 300~s of experiment as shown in Figure~\ref{gas}(i). 
\\

\begin{figure}
{{\includegraphics[width=155mm]{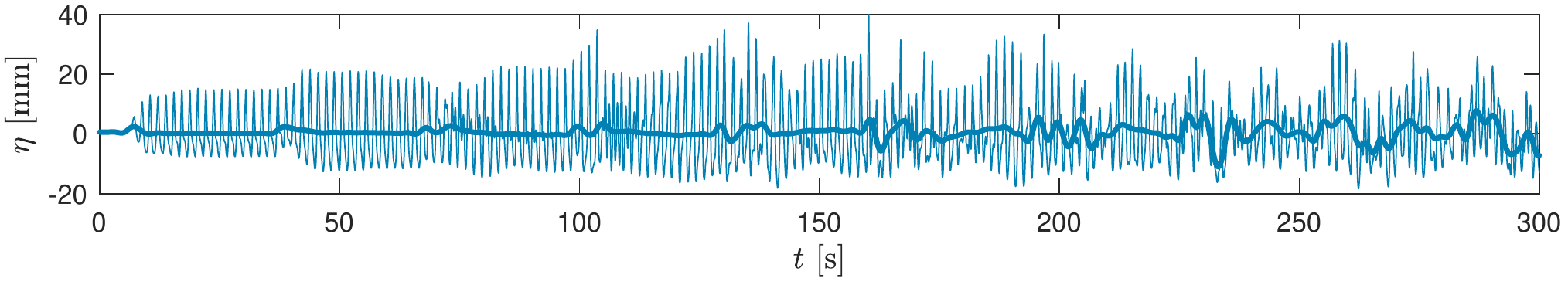}}}
\caption{ 
Time series of free surface elevation (thin line) and low-pass-filtered (thick line) recorded at the middle of the flume $x=16.89$ m.
Water depth $h=12$ cm, sinusoidal forcing with frequency $\omega/2\pi=0.6$ s$^{-1}$ and amplitude $a_0=12$ mm.
\label{eta}}
\end{figure}

A closer inspection of Figure~\ref{gas}(h, i) gives a clue for understanding the mechanisms of soliton gas outbreak. 
The energetic wave number band around $k_0$ is bleeding every $\sim 30$ s ({appearing as }vertical streaks). This period corresponds to the return time, in the field of the cameras, 
of the remnant wave front initiated at the start of the wave-maker motion. 
This periodic spectral bleeding, characterized by energy increase on a larger wave-number band, is enhanced each time the front returns.
At some point the system bifurcates to a white noise type of random wave field.
This is illustrated in Figure~\ref{eta}.
The reflected solitons that emerged from the first wave  front favorably interact with those emerging from the freshly generated sine waves. 
{This generates phase shifts that randomize the wave field after a few cycles.}
The low-pass filtered signal plotted in Figure~\ref{eta} underlines the progressive outbreak of low frequency wave components. 
\\

\begin{figure}
\hspace*{.3cm} {\bf (a)} \hspace{4.6cm} {\bf (b)} \hspace{4.6cm} {\bf (c)}\\
{{\includegraphics[height=46mm,viewport=42 183 227 355,clip]{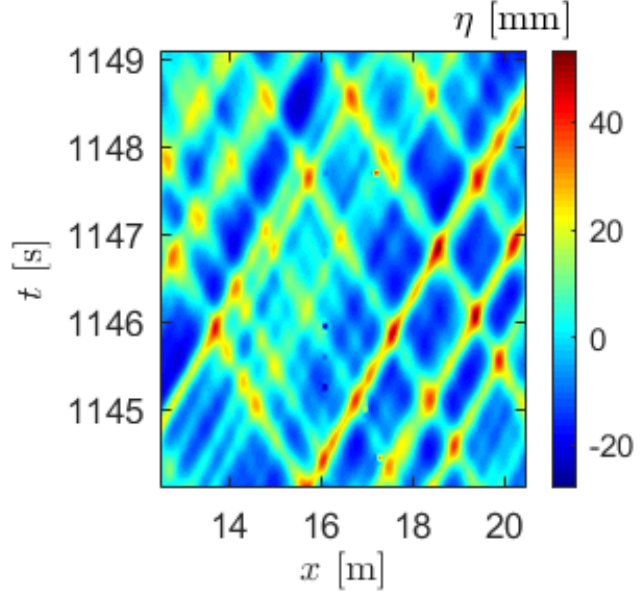}}}
\hspace{.1cm}
{{\includegraphics[height=46mm,viewport=235 8 420 180,clip]{cradon_11_1146s_18m_g.pdf}}}
\hspace{.1cm}
{{\includegraphics[height=46mm,viewport=428 8 613 180,clip]{cradon_11_1146s_18m_g.pdf}}}
\hspace*{5.7cm} {\bf (d)} \hspace{4.6cm} {\bf (e)}\\
\hspace*{5.2cm} 
{{\includegraphics[height=46mm,viewport=235 183 420 355,clip]{cradon_11_1146s_18m_g.pdf}}}
\hspace{.1cm}
{{\includegraphics[height=46mm,viewport=428 183 613 355,clip]{cradon_11_1146s_18m_g.pdf}}}
\caption{ 
(a) Close-up of a soliton gas wave field with separated (b) right-going waves and (d) left-going waves. Dashed gray lines are crests identified from local maxima (circles in c,e) in the Radon space. Corresponding crest speeds are close to $c_0$ (vertical solid lines in c,e). 
Several soliton  interactions are observed in (a), in particular a weak interaction at ($x=13.7$ m, $t=1146$~s), a strong interaction of right-going solitons centered at ($x=14.8$ m, $t=1147$~s).
\label{gascloseup}}
\end{figure}

Figure~\ref{gascloseup} provides a selected part of the wave field for the largest forcing amplitude
that emphasizes nonlinear interactions in the gas.
The local maxima in the Radon space shown in Figure~\ref{gascloseup}(c, e) indicate that the waves propagate at speeds close to $c_0$, with some random scatter.
The separation into right- and left-running waves evidences  soliton interactions. A weak interaction or head-on collision with associated phase shifts is observed at ($x=13.7$ m, $t=1146$~s). 
The same right-running soliton is overtaking a slower propagating soliton around ($x=14.8$ m, $t=1147$~s) with characteristics of a strong interaction. 
The fastest soliton is shifted ahead and the center of the interaction exhibits a two-heads feature. 
Another strong interaction of right-running solitons can be seen around ($x=13.2$ m, $t=1147.2$~s) with a quasi-simultaneous weak interaction with a left-running soliton.
For this small {(8~m, 5~s) selected region, at least 3 strong interactions and about 24} weak interactions can be identified.
This underlines that many soliton interactions occur in the gas. 
\\

\begin{figure}
\centerline
{\includegraphics[width=120mm,viewport=0 0 390 380,clip]{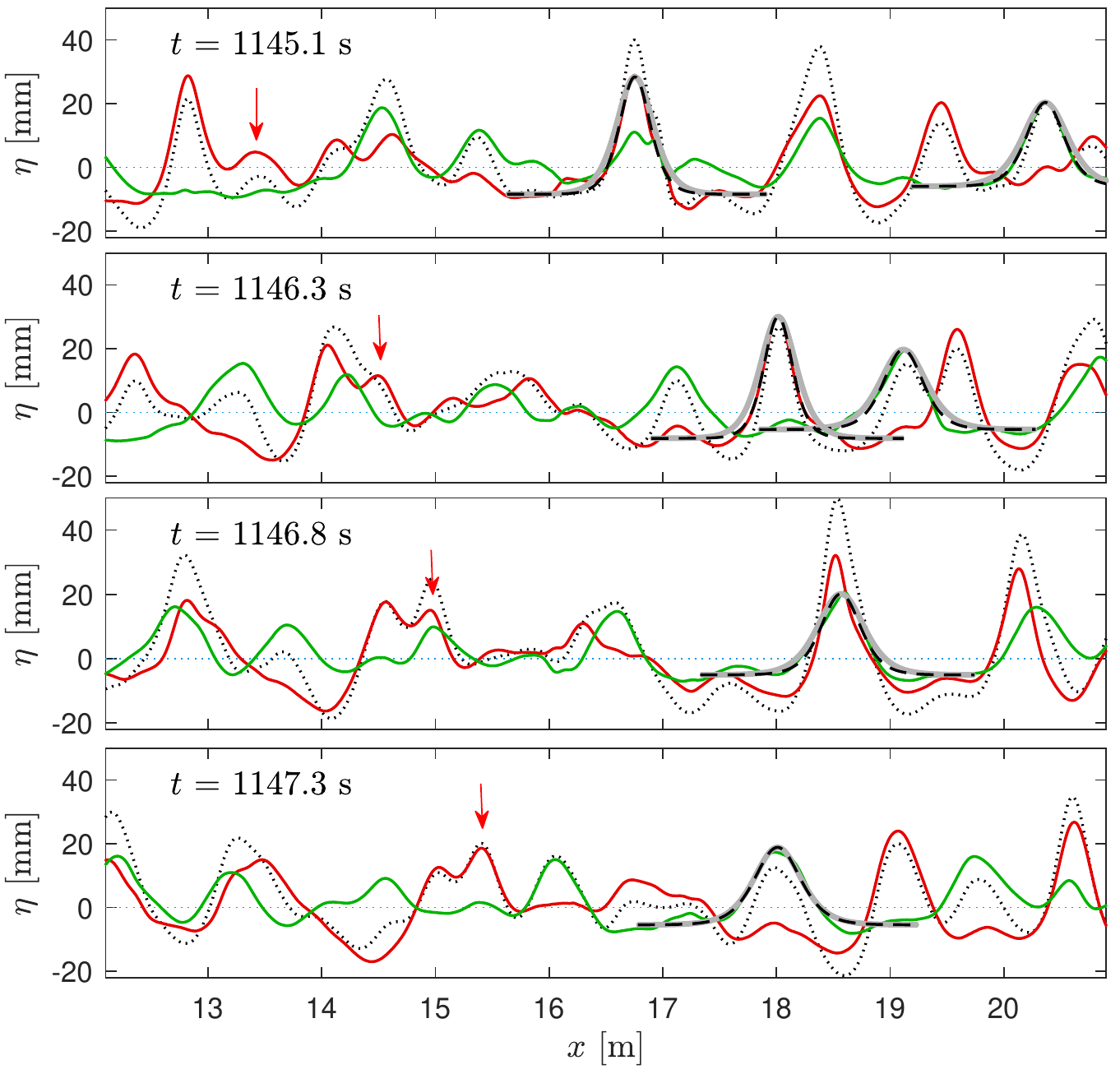}}
\caption{
Snapshots of surface elevation (dotted) separated into right-going waves (red) and left-going waves (green). 
Gray solid lines are KdV profiles, black dashed are sech$^2$ fitted profiles that closely match the KdV profiles. The arrows mark the leading soliton of the strong interaction shown in Figure~\ref{gascloseup}(b).
\label{solintfaiblefort}}
\end{figure}

The data processing presented in section~3 is applied in order to evaluate soliton characteristics in the gas.
An arbitrary criterion is defined on the standard deviation of the difference between the measured profile and a sech$^2$ fit.
This enables {us} to identify symmetric soliton-type pulses but excludes the identification of the solitons during strong interactions. 
Snapshots of surface elevation at four different times are shown in Figure~\ref{solintfaiblefort}. 
It provides an alternate view (for $x<16$~m) of the overtaking interaction described above. 
On the right-hand side of the figure, counter propagating solitons with superimposed fitted profiles are seen to collide in the third panel at ($x=18.5$ m, $t=1146.8$~s). 
Here the {detected} pulses match KdV theoretical profiles. In particular, the left-running pulse at $x=20.3$~m that just entered the range of the plot (top panel) complies to a soliton KdV shape through the interaction. 
It is noteworthy that the fitting evidences that the solitons propagate on a reference level $h_r$ that is below the quiescent water level $h$. 
\\

\begin{figure}
{
\hspace*{1.3cm} {\bf (a)} \hspace{6.6cm} {\bf (c)} \\
\hspace*{.7cm}
{{\includegraphics[width=65mm,viewport=0 0 210 295,clip]{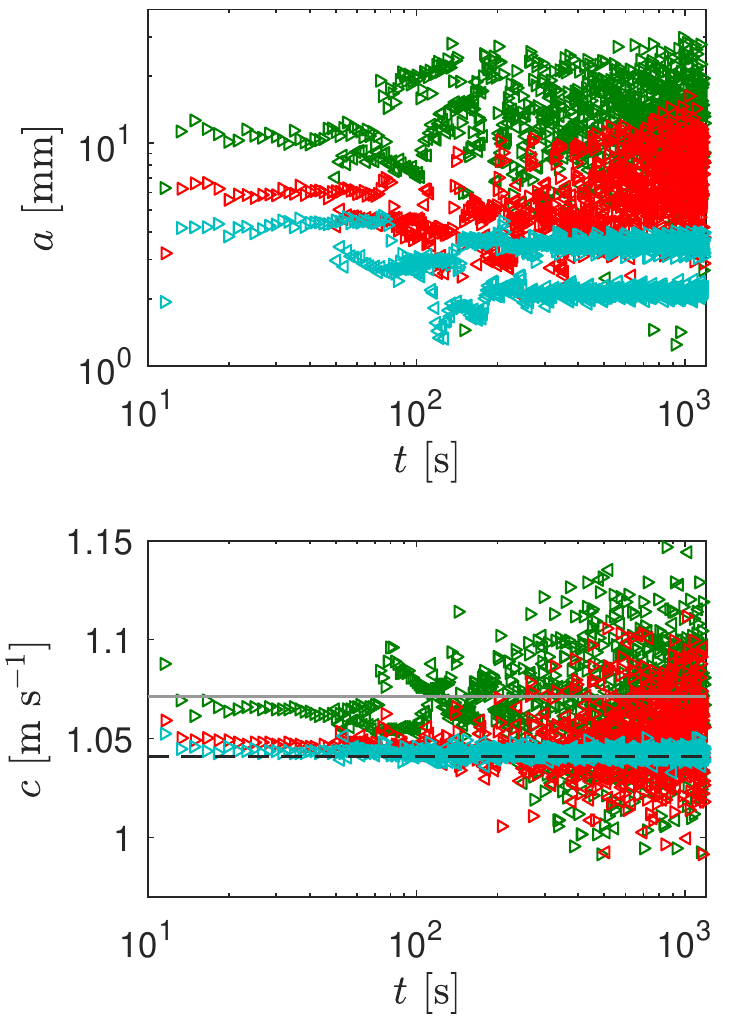}}}
\hspace{1.5cm}
{{\includegraphics[width=65mm,viewport=430 00 640 295,clip]{Cbeta_10_11_15_d.pdf}}}\\
\vspace*{-.7cm}\\ 
\hspace*{1.3cm} {\bf (b)} \hspace{6.6cm} {\bf (d)} \\
}
\caption{ 
Time variations of crest (a) maximum and (b) speed detected at $x=16.9$~m with the Radon transform.
The thin solid gray line is $c_0$, the black dashed line is $\omega/k$ fulfilling the dispersion relation (\ref{eqdisp}) with $\omega/2\pi=0.6$ s$^{-1}$.
(c) Dimensionless speed by the Radon transform against dimensionless amplitude and
(d) shape factor against dimensionless detected soliton amplitude.
Thick solid and dashed lines are for KdV and Rayleigh solutions, respectively.
Right (left) pointing triangles are for right (left) running waves. 
Water depth $h=12$ cm, sinusoidal forcing with frequency $\omega/2\pi=0.6$ s$^{-1}$ and amplitude 
$a_0=4$ mm (cyan), $a_0=6$ mm (red), $a_0=12$ mm (green).
In (d), only detected solitons with $a_s>2\,a_0$ are shown.
\label{solgaz}}
\end{figure}

The time evolution of crest maximum, detected in the middle of the flume with the Radon transform over a sliding temporal window, is plotted in Figure \ref{solgaz}(a), for three different amplitudes of the monochromatic forcing.
For times $t<50$~s, only right-running waves are present. Waves reflected by the end wall recorded thereafter have a weaker amplitude due to dissipation. 
The combination of waves reflected by the wave-maker with newly produced waves are recorded after 75~s.
Amplification in crest maximum is then observed for the largest forcing $a_0=12$~mm but the waves remain relatively organized until $t\sim 120$~s. Afterward wave crests reach various maxima, 
some exceeding 2.5 times $a_0$. Apparently randomized crest maxima is one of the soliton gas feature. 
A similar trend is obtained for $a_0=6$~mm, yet for a delayed and more progressive disorganization. In contrast, crest variability is small for the smaller forcing.
\\

The time evolution of detected crest speed is shown in Figure \ref{solgaz}b.
For the smaller amplitude of forcing, the wave speed complies to the intermediate water depth phase velocity $c=\omega/k$ with 
(\ref{eqdisp}) for the entire 20 minutes experiment 
(except for the very first waves that are at the front of the long wave produced by the departure from rest of the wave-maker), within an error range of about 5~mm~s$^{-1}$ or 0.5~\% of $c$. 
This is an indication of the accuracy of speed estimation with the Radon transform.
\\

The wave speed of the first waves is increasing with increasing forcing amplitude $a_0$. 
For the largest forcing, the right-running wave crests propagate faster than $c_0$ for $75<t<120$~s which is a clear  signature of soliton behavior. 
Subsequent disorganization characterized by a strong scattering of propagation speeds is the consequence of the multiple soliton interactions and the scattering
of solitons amplitudes.
\\

It should be emphasized that most waves propagate more slowly than $c_0$. 
Once the gas {is} formed, due to mass conservation, the solitons propagate on a reference level $h_r < h$.
\citet{osborne1986solitons} show that solitons emerging from a sinusoidal forcing should propagate at
\begin{equation}
c = c_0 \left( 1 + \frac{a_s + 3(h_r-h)}{2 h} \right) \,.
\label{cosbo}
\end{equation}
This can be also viewed as the solitons propagating against an adverse mean flow related to the trough of the initial sinusoidal wave.
In Figure~\ref{solintfaiblefort}, the right-going soliton fit gives $a/h\simeq 0.2$ and $h_r/h\simeq 0.9$ so that $a_s/h=(a+h-h_r)/h\simeq 0.3$ and $c\simeq c_0$ from (\ref{cosbo}). 
This is in good agreement with wave crest speeds detected by the Radon transform shown in Figure~\ref{solgaz}(c). 
\\

The sech$^2$ fit dimensionless shape parameter $\beta_*$ is plotted against its dimensionless soliton amplitude in Figure~\ref{solgaz}(d).
It overall follows the trend predicted by non-linear theories but with a fair amount of scattering. This is due to both strong and weak interactions.
{Detected isolated pulses are generally narrower than KdV solitons. This is probably due to the high probability for a soliton to collide with another soliton in a weak interaction, which roughly takes place every 1\,m (Figure~\ref{gascloseup}a).}
\\

As a last remark, 
doubling the forcing amplitude from 6~mm to 12~mm leads to similar scaled
soliton gas characteristics. 
{In both cases the continuous forcing provides the energy input that compensates for viscous dissipation necessary to reach a statistically quasi-steady state.}
Reducing the forcing amplitude to 4~mm leads to a completely different steady standing wave regime.
The thresholds for soliton gas occurrence might be worth further investigations.

\section{Conclusions} 

A video system is shown to accurately measure waves propagating in both directions in a constant depth flume with full reflection at both ends. From a video camera pixel resolution of 1 mm, sub-pixel interface detection leads to an
accuracy in water elevation likely better than one tenth of a millimeter when measuring waves several centimeters high. 
Crest wave speed is quantified with uncertainty of a few mm/s for waves propagating at about 1 m/s.
This is done by applying the Radon transform on the $(x, t)$ field with a relatively high resolution, 1 cm $\times$ 0.025 s providing a good compromise between accuracy and computing time. 
\\

The monochromatic forcing at one end of the flume produces wave trains that 
can degenerate into a disorganized state identified as a soliton gas.
The solitons, emerging from the periodic wave trains through non-linear steepening, experience multiple interactions that induce  various phase shifts. 
{Due to the many weak interactions (head-on collisions), the solitons detected in the gas are generally more peaky than KdV solitons.} 
They are shown to propagate onto a reference level $h_r$ located below the mean water level $h$, as a consequence of mass conservation. 
Their speed is consequently smaller than that of KdV solitons propagating at constant depth $h$.
\\

{
\citet{redorPRL} showed that both weak and strong  interactions of two solitons can be properly described with the Kaup-Boussinesq (Kaup-Broer) integrable system of equations. 
\citet{nabelek2020solutions} further suggest 
that soliton gases with many counter propagating solitons could be favorably analyzed with a dressing method for such set of equations, a promising track for future research.
Since the final states in some of our experiments are highly nonlinear with solitons dominating the dynamics, the periodic Inverse Scattering Transform for the KdV equation
\citep{Osborne2010} could also be another approach to determining the number of emerging solitons and their amplitude \citep{redor2019}.}

\section*{Acknowledgements} 

This study has received funding from the European
Research Council (ERC) under the European Union's
Horizon 2020 research and innovation programme (Grant
Agreement No. 647018-WATU). 

\bibliographystyle{spbasic}
\bibliography{mabiblio}

\end{document}